\documentclass[lettersize,journal]{IEEEtran}
\usepackage{amsmath,amsfonts,amssymb}
\usepackage{algorithmic}
\usepackage{algorithm}
\usepackage{array}
\usepackage{subcaption}
\usepackage{textcomp}
\usepackage{stfloats,flushend}
\usepackage{url}
\usepackage{hyperref}
\usepackage{verbatim}
\usepackage{bbm}
\usepackage{nicefrac,xfrac}
\usepackage{dsfont}
\usepackage{graphicx}
\usepackage[noadjust]{cite}
\newtheorem{theorem}{Theorem}
\newtheorem{corollary}{Corollary}
\newtheorem{lemma}{Lemma}
\newtheorem{identity}{Identity}
\newtheorem{definition}{Definition}
\usepackage{color}

\begin{document}

\title{\textcolor{black}{Zak-OTFS with Spread Carrier Waveforms}}
% \title{Zak-OTFS for Spread Spectrum Communication}

\author{Nishant Mehrotra$^*$, Sandesh Rao Mattu$^*$, Robert Calderbank~\IEEEmembership{Fellow,~IEEE}\vspace{-5mm}
        % <-this % stops a space
\thanks{This work is supported by the National Science Foundation under grants 2342690 and 2148212, in part by funds from federal agency and industry partners as specified in the Resilient \& Intelligent NextG Systems (RINGS) program, and in part by the Air Force Office of Scientific Research under grants FA 8750-20-2-0504 and FA 9550-23-1-0249. \\ The authors are with the Department of Electrical and Computer Engineering, Duke University, Durham, NC, 27708, USA (email: \{nishant.mehrotra,~sandesh.mattu,~robert.calderbank\}@duke.edu). \\ $*$ denotes equal contribution.\\
© 2025 IEEE. Personal use of this material is permitted. Permission from IEEE must be obtained for all other uses, in any current or future media, including reprinting/republishing this material for advertising or promotional purposes, creating new collective works, for resale or redistribution to servers or lists, or reuse of any copyrighted component of this work in other works. Published in: \textit{IEEE Wireless Communications Letters} (Early Access). DOI: \href{https://doi.org/10.1109/LWC.2025.3590254}{10.1109/LWC.2025.3590254}.}%
% This work may be submitted to the IEEE for possible publication. Copyright may be transferred without notice, after which this version may no longer be accessible.}% <-this % stops a space
%\thanks{Manuscript received April 19, 2021; revised August 16, 2021.}
}

% The paper headers
%\markboth{Journal of \LaTeX\ Class Files,~Vol.~14, No.~8, August~2021}%
%{Shell \MakeLowercase{\textit{et al.}}: A Sample Article Using IEEEtran.cls for IEEE Journals}

%\IEEEpubid{0000--0000/00\$00.00~\copyright~2021 IEEE}
% Remember, if you use this you must call \IEEEpubidadjcol in the second
% column for its text to clear the IEEEpubid mark.

\maketitle

\begin{abstract}
%An attractive feature of spread spectrum technologies such as code division multiple access (CDMA) is that it is harder to intercept or jam signals, and this feature was lost when orthogonal frequency domain modulation prevailed over CDMA in wireless standards. Legacy spread carrier waveforms are not matched to delay and Doppler shifts characteristic of 6G wireless environments, and this makes equalization very challenging. 

\textcolor{black}{Zak-OTFS (orthogonal time frequency space) modulation is a communication framework that parameterizes the wireless channel in the delay-Doppler (DD) domain, where the parameters map directly to physical attributes of the scatterers that comprise the scattering environment. As a consequence, the channel can be efficiently acquired and equalized. The Zak-OTFS carrier is a pulse in the DD domain, and the Zak transform converts it to a pulse train modulated by a tone (pulsone) in the time domain. The pulsone waveform is localized rather than spread, and it suffers from high peak-to-average power ratio (PAPR). We describe how to transform the orthonormal basis of Zak-OTFS pulsones into an orthonormal basis of spread carrier waveforms with low PAPR (only $6.58$ dB) that support communication in the presence of mobility and delay spread. This transformation is realized by a unitary transform based on the discrete affine Fourier transform. Unlike other spread modulations that achieve low PAPR by spreading information across a wider bandwidth (thus reducing the spectral efficiency), the proposed spread carrier-based Zak-OTFS achieves full spectral efficiency like pulsone-based Zak-OTFS, with $5.6$ dB lower PAPR per basis element. We demonstrate uncoded bit error rate (BER) similar to pulsone-based Zak-OTFS, and improved BER performance over competing methods based on OFDM and OTFS in high mobility \& delay spread environments.}

\end{abstract}


\section{Introduction}
\label{sec:intro}

\IEEEPARstart{Z}{ak}-OTFS (orthogonal time frequency space modulation)~\cite{bitspaper1,bitspaper2} is an attractive framework for communication, radar sensing, and integrated sensing and communications that processes signals in the delay-Doppler (DD) domain. The Zak-OTFS carrier waveform is a pulse in the DD domain, formally a quasi-periodic localized function with specific periods along delay and Doppler. When the channel delay spread is less than the delay period, and the channel Doppler spread is less than the Doppler period, the  Zak-OTFS Input/Output (I/O) relation changes slowly and predictably, and it is not subject to fading~\cite{bitspaper1,bitspaper2}. The response to a single Zak-OTFS carrier provides an image of the the scattering environment, and can be used to predict the I/O response to all other carriers. \textcolor{black}{The image of the scattering environment can support sensing in systems such as passive radars that integrate sensing and communications~\cite{bitspaper2}.}

%This property makes the Zak-OTFS carrier an ideal radar waveform that can enable integrated sensing and communication in 6G wireless~\cite{bitspaper2}.

%Zak-OTFS is realized using pulse trains modulated by tones (called pulsones)~\cite{bitspaper1,bitspaper2}, which form an orthonormal basis for DD communication. However, pulsones are difficult to realize in practice due to their high peak-to-average power ratio (PAPR)~\cite{Aug2024paper}. 

\begin{table}[!t]
    \centering
    \caption{\textcolor{black}{Comparison with prior work. $\alpha$ denotes the ratio of total vs used subcarriers in DFT-spread-OFDM ($\alpha > 1$).}}
    {\color{black}
    \setlength{\tabcolsep}{5.5pt}
    \begin{tabular}{|c|c|c|c|}
         \hline
         Method & Spectral Eff. & PAPR & Predictable? \\
         \hline
         \textbf{Spread Zak-OTFS (Ours)} & \textbf{1} & \textbf{6.6 dB} & \textbf{\checkmark} \\
         %\hline
         Pulsone Zak-OTFS~\cite{bitspaper1,bitspaper2} & 1 & 12.2 dB & \checkmark \\
         %\hline
         DFT-spread-OFDM~\cite{Wu2023_dft_s_ofdm} & $\nicefrac{1}{\alpha}$ & 6--8 dB & $\times$ \\
         OTFS~\cite{Hadani2017,Hanzo2023a,Hanzo2023b} & 1 & 10--12 dB & $\times$ \\
         %\hline
         % DFT-s-OTFS~\cite{Wu2023_dft_s_otfs} & 1 & 8 dB & $\times$ \\
         %\hline
         OFDM & 1 & 10--12 dB & $\times$ \\
         %\hline
         \hline
    \end{tabular}
    }
    \vspace*{-0.1in}
    \label{tab:prior_work}
\end{table}

The Zak-OTFS carrier is realized in the time domain as a pulse train modulated by a tone (pulsone), which suffers from high peak-to-average power ratio (PAPR)~\cite{Aug2024paper}. It is possible to reduce PAPR significantly by applying a unitary spreading filter to pulsones in the DD domain~\cite{Aug2024paper}. \textcolor{black}{We apply a unitary transform to pulsones in the \emph{time domain} to generate spread carrier waveforms with constant amplitude zero auto-correlation (CAZAC\footnote{\textcolor{black}{Examples of CAZAC sequences are Zadoff-Chu (ZC) sequences~\cite{zadoff_chu} that are used for synchronization and random access in LTE and 5G-NR.}})~\cite{Globecom2025,benedetto_phasecoded} properties. The proposed unitary transform is based on a generalization of the discrete affine Fourier transform~\cite{Ding2000frft}.} %used in signal processing

\textcolor{black}{Since the transform is unitary, the proposed method achieves full spectral efficiency ($BWT$ information symbols in $BWT$ channel uses) like pulsone-based Zak-OTFS, with $5.6$ dB lower PAPR per basis element. In contrast, existing spread modulations like DFT-spread-OFDM~\cite{Wu2023_dft_s_ofdm} achieve low PAPR by spreading information across a wider bandwidth; thus have spectral efficiency inversely proportional to the spreading factor. Table~\ref{tab:prior_work} places our contributions in the context of prior work. In Section~\ref{sec:results}, through bit error rate (BER) simulations using a six-path Vehicular-A channel model, we demonstrate uncoded $4$-QAM\footnote{\textcolor{black}{quadrature amplitude modulation}} BER similar to pulsone-based Zak-OTFS, and improved BER performance over competing methods based on OFDM, DFT-spread-OFDM~\cite{Wu2023_dft_s_ofdm} and OTFS~\cite{Hadani2017,Hanzo2023a,Hanzo2023b} in high mobility \& delay spread environments.}

%with each spread carrier waveform exhibiting $11.5$ dB lower PAPR than the corresponding pulsone.

%First, we define a discrete generalization of the fractional Fourier transform that is unitary, enabling pulsones to be mapped to spread carrier waveforms that are matched to the delay and Doppler characteristics of the wireless channel. 

%First, we show that Zak-OTFS pulsones can be transformed to time domain CAZAC sequences through a unitary transform which is a discrete generalization of the fractional Fourier transform. The unitary transformation enables defining an orthonormal basis equivalent to the pulsone basis using CAZACs. Second, we describe the end-to-end transceiver signal processing, comprising channel estimation and data demodulation, corresponding to the proposed CAZAC-based Zak-OTFS implementation. We verify our implementation through bit error rate (BER) simulations using a six-path Vehicular-A channel model, showing that the proposed CAZAC-based Zak-OTFS implementation achieves similar uncoded BER as traditional pulsone-based implementations, with each CAZAC basis element exhibiting $11.5$ dB lower PAPR over the corresponding pulsone.

\textit{Notation:} $x$ denotes a complex scalar, $\mathbf{x}$ denotes a vector with $n$th entry $\mathbf{x}[n]$, and $\mathbf{X}$ denotes a matrix with $(n,m)$th entry $\mathbf{X}[n,m]$. $(\cdot)^{\ast}$ denotes complex conjugate, $(\cdot)^{\mathsf{H}}$ denotes complex conjugate transpose and $\langle \cdot, \cdot \rangle$ denotes inner product. Calligraphic font $\mathcal{X}$ denotes operators or sets, with usage clear from context. $\emptyset$ denotes the empty set. $\mathbb{Z}$ denotes the set of integers, $\mathbb{Z}_{+}$ the set of positive integers and $\mathbb{Z}_{N}$ the set of integers modulo $N$. $(a,b)$ denotes the greatest common divisor of two integers $a,b$, $\lfloor \cdot \rfloor$ and $\lceil \cdot \rceil$ denote the floor and ceiling functions, and $(\cdot)^{-1}_N$ denotes inverse modulo $N$. $\delta[\cdot]$ denotes the Kronecker delta function, $\mathds{1}_{\{\cdot\}}$ denotes the indicator function, and $\mathbf{e}_{n}$ denotes the standard basis vector with value $1$ at location $n$ and zero elsewhere.

\section{Preliminaries: Zak-OTFS via Pulsones}
\label{sec:prelim}

Consider an $M\times N$ delay-Doppler (DD) grid\footnote{\textcolor{black}{OFDM equivalence: $M$ is number of subcarriers, $N$ is number of symbols, $\nu_p$ is subcarrier spacing. There is no cyclic prefix in Zak-OTFS.}} with $M$ delay bins (indexed by $k$) and $N$ Doppler bins (indexed by $l$), corresponding to width $\tau_p$ along delay and width $\nu_p$ along Doppler with $\tau_p \nu_p = 1$. The basis element corresponding to the $(k_0,l_0)$th DD bin, where $0\leq k_0 \leq (M-1),~0\leq l_0 \leq (N-1)$, is $\delta[k-k_0]\delta[l-~l_0]$. The corresponding discrete time realization\footnote{\textcolor{black}{Superscript $(\cdot)^{(\mathrm{p})}$ denotes pulsone.}} via the inverse discrete Zak transform is~\cite{dzt}:
\begin{equation}
    \label{eq:zakotfs1}
    \mathbf{x}^{(\mathrm{p})}_{(k_0,l_0)}[n] = \frac{1}{\sqrt{N}} \sum_{d \in \mathbb{Z}} e^{\frac{j2\pi}{N} d l_0} \delta[n-k_0-dM],
\end{equation}
which is termed \emph{pulsone} in the Zak-OTFS literature~\cite{bitspaper1,bitspaper2}. The pulsone basis is orthonormal, and $MN$ information symbols can be transmitted by modulating different pulsones:
\begin{equation}
    \label{eq:zakotfs2}
    \mathbf{x}_{t}^{(\mathrm{p})}[n] = \sum_{k_0=0}^{M-1} \sum_{l_0=0}^{N-1} \mathbf{X}[k_0,l_0] \mathbf{x}^{(\mathrm{p})}_{(k_0,l_0)}[n].
\end{equation}

%Each DD bin can be represented by the $(k+lM)$th standard basis vector $\mathbf{e}_{k+lM} \in \{0,1\}^{MN\times 1}$.
%For example, the DD bin corresponding to $(k,l) = (0,0)$ is the vector $\mathbf{e}_{0,0} = \begin{bmatrix}
%    1 & 0 & \cdots & 0
%\end{bmatrix}^{\mathsf{T}}$. 

\section{Spread Carrier Waveforms in Zak-OTFS} %Mapping Pulsone Basis to TD CAZACs
\label{sec:cazac_basis}

A drawback of the pulsone basis in~\eqref{eq:zakotfs1} is its high PAPR~\cite{Aug2024paper}. In this Section, we design spread carrier waveforms within the Zak-OTFS framework by mapping the pulsone basis to CAZACs via a unitary transform (Definition~\ref{def:frft}). In the sequel, we make use of the following two identities extensively.

\begin{identity}[\cite{murty2017evaluation}]
    \label{idty:sumrootsofunity}
    The sum of all $N$th roots of unity satisfies:
    \begin{align*}
        \sum_{n=0}^{N-1}e^{\frac{j2\pi}{N}kn} = \begin{cases}
        N \quad \text{if } \ k \equiv 0 \bmod (N) \\
        0 \quad \ \text{otherwise}
        \end{cases}.
    \end{align*}
\end{identity}

\begin{identity}[\cite{murty2017evaluation}]
    \label{idty:gauss_sum}
    The quadratic Gauss sum is given by:
    \begin{align*}
        \sum_{n=0}^{N-1} e^{\frac{j2\pi}{N}(an^2+bn)} = \epsilon_{N}\sqrt{N}\left(\frac{a}{N}\right)_J e^{-\frac{j2\pi}{N}(4a)^{-1}_{N} b^2}, 
    \end{align*}
where $N$ is odd, $(a, N) = 1$, $a, b \in \mathbb{Z}_+$, $\epsilon_{N} = 1$ if $N \equiv 1 \bmod 4$ and $\epsilon_{N} = j$ if $N \equiv 3 \bmod 4$. $\left(\frac{a}{b}\right)_J = \prod_{i=1}^{k} \left(\frac{a}{p_i}\right)_L^{\alpha_i}$ denotes the Jacobi symbol, which for any integer $a$ and positive odd integer $b$, is the product of the Legendre symbols corresponding to the prime factors of $b = \prod_{i = 1}^{k} p_i^{\alpha_i}$, where, for integers $a$ and odd primes $p$:
\begin{align*}
    \left(\frac{a}{p}\right)_L = \begin{cases}
            0 \ ~~\text{if $a \equiv 0 \bmod p$} \\
            1 \ ~~\text{if $a \not\equiv 0 \bmod p$, $\exists x \in \mathbb{Z}:a \equiv x^2 \bmod p$} \\
            -1~\text{if $a \not\equiv 0 \bmod p$, $\not\exists x \in \mathbb{Z}:a \equiv x^2 \bmod p$}
        \end{cases}.
\end{align*}
    % \begin{align*}
    %     \epsilon_{N} = \begin{cases}
    %         1~~~\text{if $N \equiv 1 \bmod 4$} \\
    %         j~~~\text{if $N \equiv 3 \bmod 4$}
    %     \end{cases},
    % \end{align*}
% and $\left(\frac{a}{b}\right)_J$ denotes the Jacobi symbol, which for any integer $a$ and positive odd integer $b$, is the product of the Legendre symbols corresponding to the prime factors of $b = \prod_{i = 1}^{k} p_i^{\alpha_i}$:
% \begin{align*}
%     \left(\frac{a}{b}\right)_J = \prod_{i=1}^{k} \left(\frac{a}{p_i}\right)_L^{\alpha_i},
% \end{align*}
% where, for integers $a$ and odd primes $p$, $\left(\frac{a}{p}\right)_L$ is:
% \begin{align*}
%     \left(\frac{a}{p}\right)_L = \begin{cases}
%             0 \ ~~\text{if $a \equiv 0 \bmod p$} \\
%             1 \ ~~\text{if $a \not\equiv 0 \bmod p$, $\exists x \in \mathbb{Z}:a \equiv x^2 \bmod p$} \\
%             -1~\text{if $a \not\equiv 0 \bmod p$, $\not\exists x \in \mathbb{Z}:a \equiv x^2 \bmod p$}
%         \end{cases}.
% \end{align*}
\end{identity}

\begin{definition}
    \label{def:frft}
    The generalized discrete affine Fourier transform (GDAFT) of an $MN$-length sequence $\mathbf{x}$ is given by:
    \begin{align*}
        \mathcal{F}_a\mathbf{x}[n] = \frac{1}{\sqrt{MN}} \sum_{m=0}^{MN-1} e^{\frac{j2\pi}{MN} (An^2+Bnm+Cm^2)} \mathbf{x}[m],
    \end{align*}
    where $n\in\{0,\cdots,MN-1\}$, $A,B,C$ are co-prime to $MN$.
\end{definition}

\textit{Note:} The GDAFT is similar in form to the traditional discrete affine Fourier transform~\cite{Ding2000frft} used in signal processing.

\begin{lemma}
    \label{lmm:frft_unitary}
    The GDAFT in Definition~\ref{def:frft} is a unitary transform. Therefore, the inverse GDAFT is given by:
    \begin{align*}
        \mathcal{F}_a^{-1}\mathbf{x}[m] = \frac{1}{\sqrt{MN}} \sum_{n=0}^{MN-1} e^{-\frac{j2\pi}{MN} (An^2+Bnm+Cm^2)} \mathbf{x}[n].
    \end{align*}
\end{lemma}
\begin{IEEEproof}
    We may represent the GDAFT by an ${MN\times MN}$ matrix $\mathbf{U}$, with the $(n,m)$th element of $\mathbf{U}$ given by $\mathbf{U}[n,m] = \frac{1}{\sqrt{MN}} e^{\frac{j2\pi}{MN} (An^2+Bnm+Cm^2)}$. Therefore, $\mathbf{U}^{\mathsf{H}}\mathbf{U}$ is given by:
    \begin{align*}
        (\mathbf{U}^{\mathsf{H}}\mathbf{U})[m_1,m_2] &=\!\frac{1}{MN}\!\sum_{n = 0}^{MN-1}\!e^{\frac{j2\pi}{MN} (Bn+C(m_2+m_1))(m_2-m_1)} \nonumber \\
        &= \delta[m_1 = m_2],
    \end{align*}
    where the final expression follows from Identity~\ref{idty:sumrootsofunity} and the fact that $0 \leq m_1,m_2 \leq (MN-1)$. Hence, $\mathbf{U}^{\mathsf{H}}\mathbf{U}$ is the $MN\times MN$ identity matrix, i.e., the GDAFT is a unitary transform.
\end{IEEEproof}

\begin{theorem}
    \label{thm:frft_pp_tdcazac}
    The GDAFT in Definition~\ref{def:frft} maps the discrete time point pulsone given by~\eqref{eq:zakotfs1} localized in the discrete DD domain at $(k_0,l_0)$ to a spread carrier CAZAC sequence\footnote{\textcolor{black}{Superscript $(\cdot)^{(\mathrm{c})}$ denotes spread carrier.}}:
    \begin{align*}
        \mathbf{x}^{(\mathrm{c})}_{(k_0,l_0)}[n] = \mathcal{F}_a\mathbf{x}_{p}[n] = &\frac{e^{\frac{j2\pi}{MN} (An^2+Bnk_0 +Ck_0^2)}}{\sqrt{MN}} \epsilon_N \left(\frac{CM}{N}\right)_J \nonumber \\ 
        & \times e^{-\frac{j2\pi}{N} (4CM)_N^{-1} (Bn + l_0 + 2Ck_0)^2}.
    \end{align*}
    % \begin{align*}
    %     \mathbf{x}[m] = \sum_{d \in \mathbb{Z}} e^{\frac{j2\pi}{N} d l_0} \delta[m-k_0-dM],
    % \end{align*}
\end{theorem}
\begin{IEEEproof}
    Substituting the point pulsone from~\eqref{eq:zakotfs1} in Definition~\ref{def:frft}, and defining $\mathbf{x}^{(\mathrm{c})}_{(k_0,l_0)}[n] = \mathcal{F}_a\mathbf{x}^{(\mathrm{p})}_{(k_0,l_0)}[n]$,
\begin{align*}
    \mathbf{x}^{(\mathrm{c})}_{(k_0,l_0)}[n] = \frac{1}{N\sqrt{M}} \sum_{m=0}^{MN-1} \sum_{d \in \mathbb{Z}} &e^{\frac{j2\pi}{MN} (An^2+Bnm+Cm^2)} e^{\frac{j2\pi}{N} d l_0} \nonumber \\ &\times \delta[m-k_0-dM].
\end{align*}

The delta function takes value $1$ when $m = k_0 + dM$, for all $d \in \{0,\cdots,N-1\}$. Therefore,
\begin{align*}
    \mathbf{x}^{(\mathrm{c})}_{(k_0,l_0)}[n] & \nonumber \\
    %&\hspace{-10mm}=\!\frac{1}{N\sqrt{M}}\!\sum_{d = 0}^{N-1}\!e^{\frac{j2\pi}{MN} (An^2+Bn(k_0 + dM)+C(k_0 + dM)^2+ d M l_0)} \nonumber \\
    &\hspace{-10mm}=\!\frac{e^{\frac{j2\pi}{MN} (An^2+Bnk_0 +Ck_0^2)}}{N\sqrt{M}}\!\sum_{d = 0}^{N-1}\!e^{\frac{j2\pi}{N} (CMd^2+d(Bn + l_0 + 2Ck_0))}. % \nonumber \\
    %&= \frac{e^{\frac{j2\pi}{MN} (An^2+Bnk_0 +Ck_0^2)}}{\sqrt{M}} \epsilon_N \left(\frac{CM}{N}\right)_J e^{-\frac{j2\pi}{N} (4CM)_N^{-1} (Bn + l_0 + 2Ck_0)^2},
\end{align*}
From Identity~\ref{idty:gauss_sum}, we obtain the expression in Theorem~\ref{thm:frft_pp_tdcazac}.
\end{IEEEproof}

\textit{Note:} The expression for $\mathbf{x}^{(\mathrm{c})}_{(k_0,l_0)}[n]$ in Theorem~\ref{thm:frft_pp_tdcazac} is equivalent to a (scaled) generalized Zadoff-Chu sequence~\cite{Srdjan2010,Globecom2025} (which is a member of the CAZAC family) of the form:
\begin{equation*}
    \mathbf{x}[n] = e^{\frac{j2\pi}{MN}\big(u\frac{n(n+1)}{2} + kn + \gamma\big)},
\end{equation*}
with parameters $u = 2(A-(4C)^{-1}_{N}B^2)$, $k = (4C)^{-1}_{N}(B^2-2Bl_0)-A$, $\gamma = -(k_0l_0+(4C)^{-1}_{N}l_0^2)$.

\begin{corollary}
    \label{corr:frft_pp_tdcazac}
    Theorem~\ref{thm:frft_pp_tdcazac} equivalently states that the GDAFT in Definition~\ref{def:frft} maps the orthonormal pulsone basis to another orthonormal spread carrier CAZAC basis, with both bases parameterized by $(k_0,l_0),~0\leq k_0 \leq (M-1),~0\leq l_0 \leq (N-1)$.
\end{corollary}
\begin{IEEEproof}
    From Lemma~\ref{lmm:frft_unitary}, we know that the GDAFT is a unitary transform. Therefore, the GDAFT maps orthonormal bases to orthonormal bases and preserves inner products. Since pulsones form an orthonormal basis~\cite{bitspaper1,bitspaper2}, the GDAFT maps the pulsone basis to another orthonormal basis formed by CAZACs of the form given in Theorem~\ref{thm:frft_pp_tdcazac}.
\end{IEEEproof}

\section{Time Domain Spread Carrier Processing} %spread carrier-based Zak-OTFS Processing
\label{sec:cazac_proc}

\subsection{Mounting Information Symbols on Spread Carriers}
\label{subsec:cazac_proc_infosym}

From Theorem~\ref{thm:frft_pp_tdcazac}, we observe that analogous to~\eqref{eq:zakotfs2}, we can transmit $MN$ independent information symbols by modulating the spread carrier basis formed by the CAZACs:
\begin{equation}
    \label{eq:cazacproc1}
    \mathbf{x}_{t}^{(\mathrm{c})}[n] = \sum_{k_0=0}^{M-1} \sum_{l_0=0}^{N-1} \mathbf{X}[k_0,l_0] \mathbf{x}^{(\mathrm{c})}_{(k_0,l_0)}[n].
\end{equation}
% \begin{align*}
%     %\label{eq:cdma1}
%     \mathbf{x}_{t}[n] = &\frac{1}{\sqrt{MN}} \epsilon_N \left(\frac{CM}{N}\right)_J \sum_{k_0=0}^{M-1} \sum_{l_0=0}^{N-1} \mathbf{X}[k_0,l_0] \nonumber \\ & \times e^{\frac{j2\pi}{MN} (An^2+Bnk_0 +Ck_0^2)} \nonumber \\ & \times e^{-\frac{j2\pi}{N} (4CM)_N^{-1} (Bn + l_0 + 2Ck_0)^2}.
% \end{align*}

\subsection{Time Domain System Model}
\label{subsec:cazac_proc_sysmodel}

We describe a general system model applicable to both pulsone-based and spread carrier-based Zak-OTFS. After mounting the information symbols on either basis, the time domain signal is transmitted through an \emph{effective channel}\footnote{\textcolor{black}{The effective channel approximates the physical channel when all paths are resolvable in delay with bandwidth $BW$ and in Doppler with time $T$.}} which encompasses the effects of the linear time-varying physical channel and transmit \& receive pulse shaping \cite[Eq. (7)]{bitspaper2}. The received time domain signal is given by:
%parameterized by $(h_i, \tau_i, \nu_i)$ tuples which correspond to the channel gain, delay, and Doppler of path $i$ and $i=0, 1, \cdots, P-1$, where $P$ is the number of paths. implementations via the pulsone basis per \eqref{eq:zakotfs2} and the CAZAC basis per \eqref{eq:cazacproc1}
\begin{align}
    \mathbf{y}[n] = \sum_{k_0 = 0}^{M-1} \sum_{l_0 = 0}^{N-1} &\sum_{k, l \in \mathbb{Z}}\mathbf{h}_{\mathrm{eff}}[k, l] \mathbf{x}_{(k_0, l_0)}[n-k]e^{\frac{j2\pi}{MN}l(n-k)} \nonumber \\ 
    &~~~\times\mathbf{X}[k_0, l_0] + \mathbf{w}[n],
    \label{eq:sys_model1}
\end{align}
where $\mathbf{x}_{(k_0, l_0)}[n]$ is either the $(k_0,l_0)$th pulsone from~\eqref{eq:zakotfs1} or the CAZAC from Theorem~\ref{thm:frft_pp_tdcazac}, $\mathbf{w}[n]$ denotes additive Gaussian noise, and $\mathbf{h}_{\mathrm{eff}}[k, l]$ represents the effective channel. 

Rewriting \eqref{eq:sys_model1} in matrix-vector form:
\begin{align}
    \label{eq:sys_model2}
    \mathbf{y} = \mathbf{H}\mathbf{x} + \mathbf{w},
\end{align}
where $\mathbf{y}$ is the $MN \times 1$ received signal vector, $\mathbf{x}$ is the $MN \times 1$ transmit signal vector with ${\mathbf{x}[k_0+l_0M] = \mathbf{X}[k_0, l_0]}$, $\mathbf{H}$ is the $MN \times MN$ channel matrix with $\mathbf{H}[n, k_0+l_0M] = \sum_{k, l \in \mathbb{Z}}\mathbf{h}_{\mathrm{eff}}[k, l] \mathbf{x}_{(k_0, l_0)}{[n-k]}e^{\frac{j2\pi}{MN}l(n-k)}$, and $\mathbf{w} \sim \mathcal{CN}(\mathbf{0},\sigma^2\mathbf{I})$ is the $MN \times 1$ vector of additive complex Gaussian noise at the receiver.

We assume transmission occurs in two stages - (i) a \emph{pilot stage}, where a known (pilot) signal is transmitted to enable the receiver to estimate the channel, and (ii) a \emph{data stage} where unknown data signals and transmitted, with symbol detection performed at the receiver based on channel estimates from the pilot stage. Therefore, the combined system model across the two stages can be written as:
\begin{equation}
    \label{eq:sys_model3}
    \begin{bmatrix}
        \mathbf{y}_{p} & \mathbf{y}_{d}
    \end{bmatrix} = \mathbf{H}\begin{bmatrix}
        \mathbf{x}_{p} & \mathbf{x}_{d}
    \end{bmatrix} + \begin{bmatrix}
        \mathbf{w}_{p} & \mathbf{w}_{d}
    \end{bmatrix},
\end{equation}
where we have assumed that the channel remains constant across the two stages. For simplicity, we henceforth assume the pilot signal is the standard basis vector $\mathbf{x}_{p} = \mathbf{e}_{(k_p+l_p M)}$, for some $(k_p,l_p),~0\leq k_p \leq (M-1),~0\leq l_p \leq (N-1)$.

%At the receiver, the channel is equalized using the channel matrix $\mathbf{H}$ and symbol estimates are obtained through minimum distance detection on the equalized symbols. 

\subsection{Channel Estimation (Pilot Stage)}
\label{subsec:cazac_proc_chest}

In the pilot stage, the receiver estimates the effective channel $\mathbf{h}_{\mathrm{eff}}[k, l]$ by solving the maximum likelihood problem:
\begin{align}
    \label{eq:chest1}
    \min_{\mathbf{h}_{\mathrm{eff}}[k, l]} \sum_{n = 0}^{MN-1} \bigg| \mathbf{y}_{p}[n] - \sum_{k, l \in \mathbb{Z}}&\mathbf{h}_{\mathrm{eff}}[k, l] \mathbf{x}_{(k_p, l_p)}[n-k] \nonumber \\ &\times e^{\frac{j2\pi}{MN}l(n-k)} \bigg|^{2},
\end{align}
and forms an estimate of the matrix $\mathbf{H}$ in~\eqref{eq:sys_model3} as $\widehat{\mathbf{H}}[n, k_0+l_0M] = \sum_{k, l \in \mathbb{Z}}\widehat{\mathbf{h}}_{\mathrm{eff}}[k, l] \mathbf{x}_{(k_0, l_0)}{[n-k]}e^{\frac{j2\pi}{MN}l(n-k)}$.

It was shown in~\cite{Aug2024paper} that the solution to~\eqref{eq:chest1} is the \emph{cross-ambiguity function} of the received and transmitted signals:
\begin{equation}
    \label{eq:chest2}
    \widehat{\mathbf{h}}_{\mathrm{eff}}[k, l] = \mathbf{A}_{\mathbf{y}_{p},\mathbf{x}_{(k_p, l_p)}}[k, l].
\end{equation}

\begin{definition}[\cite{benedetto_phasecoded}]
    \label{def:tdamb}
    The \emph{cross-ambiguity function} of two periodic sequences $\mathbf{x}[n]$ and $\mathbf{y}[n]$ of period $MN$ is given by:
    \begin{align}
        \mathbf{A}_{\mathbf{y},\mathbf{x}}[k, l] &= \frac{1}{MN}\sum_{n=0}^{MN-1}\mathbf{y}[n]\mathbf{x}^*[n-k]e^{-\frac{j2\pi}{MN}l(n-k)}.
    \end{align}

    When $\mathbf{x}[n] = \mathbf{y}[n]$, $\mathbf{A}_{\mathbf{x},\mathbf{x}}[k, l]$ is called the \emph{self-ambiguity function} of $\mathbf{x}[n]$, which is written as $\mathbf{A}_{\mathbf{x}}[k, l]$ for brevity.
\end{definition}

\begin{theorem}
    \label{thm:frft_amb}
    The cross-ambiguity function of the GDAFT per Definition~\ref{def:frft} of two sequences $\mathbf{x}[n]$ and $\mathbf{y}[n]$ is given by:
    \begin{align*}
        \mathbf{A}_{\mathcal{F}_a\mathbf{y},\mathcal{F}_a\mathbf{x}}[k, l] &= e^{\frac{j2\pi}{MN} (-Ak^2+lk+CB^{-2}_{{}_{MN}} (2Ak-l)^2)} \mathbf{A}_{\mathbf{y},\mathbf{x}}[\bar{k}, \bar{l}],
    \end{align*}
    where $\bar{k} = -B^{-1}_{{}_{MN}} (2Ak-l)$, $\bar{l} = 2CB^{-1}_{{}_{MN}} (2Ak-l) - Bk$.
\end{theorem}
\begin{IEEEproof}
We may express the cross-ambiguity function as the inner product of two TD sequences:
\begin{align*}
    %\label{eq:amb1}
    \mathbf{A}_{\mathbf{y},\mathbf{x}}[k, l] &= \frac{1}{MN} \bigg\langle \mathbf{y}, \mathcal{D}_{k,l} \mathbf{x} \bigg \rangle, \\
    \label{eq:amb2}
    (\mathcal{D}_{k,l} \mathbf{x})[n] &= \mathbf{x}[n-k]e^{\frac{j2\pi}{MN}l(n-k)}.
\end{align*}

Let $\mathcal{U}$ and $\mathcal{U}^{\mathsf{H}}$ denote the (unitary) operators corresponding to the GDAFT and inverse GDAFT from Definition~\ref{def:frft} and Lemma~\ref{lmm:frft_unitary} respectively, such that $\mathcal{F}_a\mathbf{x}[n]  = (\mathcal{U}\mathbf{x})[n]$ and $\mathcal{F}_a^{-1}\mathbf{x}[m]  = (\mathcal{U}^{\mathsf{H}}\mathbf{x})[m]$. Therefore, the cross-ambiguity of the GDAFT of $\mathbf{y}$ and the GDAFT of $\mathbf{x}$ is:
\begin{align*}
    %\label{eq:amb3}
    \mathbf{A}_{\mathcal{F}_a\mathbf{y},\mathcal{F}_a\mathbf{x}}[k, l] &= \frac{1}{MN} \bigg\langle \mathcal{U}\mathbf{y}, \mathcal{D}_{k,l} \mathcal{U}\mathbf{x} \bigg \rangle \nonumber \\ 
    &= \frac{1}{MN} \bigg\langle \mathbf{y}, \mathcal{U}^{\mathsf{H}}\mathcal{D}_{k,l} \mathcal{U}\mathbf{x} \bigg \rangle,
\end{align*}
where the final expression follows from the fact that the operator $\mathcal{U}$ is unitary. Note that by definition:
\begin{align*}
    %\label{eq:amb4}
    (\mathcal{D}_{k,l} \mathcal{U}\mathbf{x})[n] = \frac{1}{\sqrt{MN}} &\sum_{m=0}^{MN-1} e^{\frac{j2\pi}{MN} (A(n-k)^2+B(n-k)m)} \nonumber \\ 
    &~\qquad \times e^{\frac{j2\pi}{MN} (Cm^2 + l(n-k))} \mathbf{x}[m].
\end{align*}

Therefore,
\begin{align*}
    %\label{eq:amb5}
    (\mathcal{U}^{\mathsf{H}}\mathcal{D}_{k,l} \mathcal{U}\mathbf{x})[m]\!&=\!\frac{1}{MN} \sum_{n=0}^{MN-1} e^{-\frac{j2\pi}{MN} (An^2+Bnm+Cm^2 - l(n-k))} \nonumber \\ 
    &\times \sum_{m'=0}^{MN-1} e^{\frac{j2\pi}{MN} (A(n-k)^2+B(n-k)m'+Cm'^2 )} \mathbf{x}[m'] \nonumber \\
    %&= \frac{1}{MN} \sum_{n=0}^{MN-1} e^{\frac{j2\pi}{MN} (A(n-k)^2-An^2-Bnm-Cm^2)} \nonumber \\ 
    %&\times \sum_{m'=0}^{MN-1} e^{\frac{j2\pi}{MN} (B(n-k)m'+Cm'^2 + l(n-k))} \mathbf{x}[m'] \nonumber \\
    %&= \frac{1}{MN} \sum_{n=0}^{MN-1} e^{\frac{j2\pi}{MN} (Ak^2-2Akn-Bnm-Cm^2+l(n-k))} \sum_{m'=0}^{MN-1} e^{\frac{j2\pi}{MN} (B(n-k)m'+Cm'^2)} \mathbf{x}[m'] \nonumber \\
    &= \frac{1}{MN} \sum_{m'=0}^{MN-1} \mathbf{x}[m'] e^{\frac{j2\pi}{MN} (Ak^2-Cm^2-lk-Bkm')} \nonumber \\ 
    &\times\!e^{\frac{j2\pi}{MN} Cm'^2}\!\sum_{n=0}^{MN-1}\!e^{\frac{j2\pi}{MN} ((Bm'+l-2Ak-Bm) n)}.
\end{align*}

From Identity~\ref{idty:sumrootsofunity}, the inner summation over $n$ vanishes unless $(Bm'+l-2Ak-Bm) \equiv 0 \bmod (MN)$, when it takes value $MN$. Therefore, $Bm' = Bm + (2Ak-l) + fMN \implies m' = m + B^{-1}_{{}_{MN}} (2Ak-l) + B^{-1}_{{}_{MN}}fMN$ for $f \in \mathbb{Z}$. Therefore:
\begin{align*}
    %\label{eq:amb6}
    (\mathcal{U}^{\mathsf{H}}\mathcal{D}_{k,l} \mathcal{U}\mathbf{x})[m]\!&=\!\mathbf{x}[m + B^{-1}_{{}_{MN}} (2Ak-l) + B^{-1}_{{}_{MN}}fMN] \nonumber \\ &\times e^{\frac{j2\pi}{MN} (Ak^2-Cm^2-lk-k(Bm + (2Ak-l) + fMN))} \nonumber \\ &\times e^{\frac{j2\pi}{MN} (C(m + B^{-1}_{{}_{MN}} (2Ak-l) + B^{-1}_{{}_{MN}}fMN)^2)} \nonumber \\
    &= \mathbf{x}[m + B^{-1}_{{}_{MN}} (2Ak-l)] e^{\frac{j2\pi}{MN} (Ak^2-Cm^2-lk)} \nonumber \\
    &e^{\frac{j2\pi}{MN} (-k(Bm + (2Ak-l)) + C(m + B^{-1}_{{}_{MN}} (2Ak-l))^2)},
\end{align*}
where the final expression follows from the $MN$-periodicity of $\mathbf{x}[n]$ and the $MN$-th roots of unity. On substituting we obtain the desired result. %expression in the Theorem statement.
%\begin{align*}
    %%\label{eq:amb7}
    %\mathbf{A}_{\mathcal{F}_a\mathbf{y},\mathcal{F}_a\mathbf{x}}[k, l] &= e^{\frac{j2\pi}{MN} (-Ak^2+lk+CB^{-2}_{{}_{MN}} (2Ak-l)^2)} \mathbf{A}_{\mathbf{y},\mathbf{x}}[\bar{k}, \bar{l}],
%    %\frac{1}{MN} \sum_{m = 0}^{MN-1} \mathbf{x}[m] \mathbf{x}^{*}[m + B^{-1}_{{}_{MN}} (2Ak-l)] e^{-\frac{j2\pi}{MN} (Ak^2-Cm^2-lk-Bk(m + B^{-1}_{{}_{MN}} (2Ak-l)) + C(m + B^{-1}_{{}_{MN}} (2Ak-l))^2)} \nonumber \\
%    %&= \frac{1}{MN} \sum_{m = 0}^{MN-1} \mathbf{x}[m] \mathbf{x}^{*}[m + B^{-1}_{{}_{MN}} (2Ak-l)] e^{-\frac{j2\pi}{MN} (Ak^2-lk - (2Ak-l)k + CB^{-2} (2Ak-l)^2 + (2CB^{-1}_{{}_{MN}} (2Ak-l) - Bk) m)} \nonumber \\
%    %&= \frac{e^{\frac{j2\pi}{MN} (-Ak^2+lk+CB^{-2} (2Ak-l)^2)}}{MN} \sum_{m = 0}^{MN-1} \mathbf{x}[m] \mathbf{x}^{*}[m + B^{-1}_{{}_{MN}} (2Ak-l)] e^{-\frac{j2\pi}{MN} (2CB^{-1}_{{}_{MN}} (2Ak-l) - Bk) (m + B^{-1}_{{}_{MN}} (2Ak-l))} \nonumber \\
%\end{align*}
%where
%\begin{align*}
%    %\label{eq:amb8}
%    \bar{k} &= -B^{-1}_{{}_{MN}} (2Ak-l), \nonumber \\
%    \bar{l} &= 2CB^{-1}_{{}_{MN}} (2Ak-l) - Bk.
%\end{align*}
\end{IEEEproof}

\begin{lemma}
    \label{lmm:heff_cryst}
    The channel estimate via~\eqref{eq:chest2} equals $\mathbf{h}_{\mathrm{eff}}[k, l]$ when the \emph{channel support}, $\mathcal{S} = \{(k,l):\mathbf{h}_{\mathrm{eff}}[k, l] \neq 0\}$, satisfies the \emph{crystallization condition}:
    \begin{equation*}
        \mathcal{S} \cap \bigg(\bigcup_{n,m \neq (0,0)} (\mathcal{S} + (k'_{n,m},l'_{n,m})) \bigg) = \emptyset,
    \end{equation*}
    where, for spread carrier-based Zak-OTFS with GDAFT parameters $A,B,C$, $k'_{n,m} = -2CB^{-1}_{{}_{MN}} nM - B^{-1}_{{}_{MN}} mN$, $l'_{n,m} = (B-4ACB^{-1}_{{}_{MN}}) nM - B^{-1}_{{}_{MN}} 2A mN$. For pulsone-based Zak-OTFS, $k'_{n,m} = nM$, $l'_{n,m} = mN$.
\end{lemma}
\begin{IEEEproof}
    By definition, in the absence of noise, we have:
    \begin{align*}
        \widehat{\mathbf{h}}_{\mathrm{eff}}[k', l'] &= \mathbf{A}_{\mathbf{y}_{p},\mathbf{x}_{(k_p, l_p)}}[k', l'] \nonumber \\
        &=\!\sum_{k, l \in \mathbb{Z}}\!\mathbf{h}_{\mathrm{eff}}[k, l] \mathbf{A}_{\mathbf{x}_{(k_p, l_p)}}[k'-k, l'-l] e^{\frac{j2\pi}{MN}l(k'-k)}.
    \end{align*}

    For pulsone-based Zak-OTFS, we know from~\cite{Aug2024paper} that:
    \begin{equation*}
        \mathbf{A}_{\mathbf{x}_{(k_p, l_p)}^{(\mathrm{p})}}[k, l] = \sum_{n,m\in\mathbb{Z}} e^{\frac{j2\pi}{N}nl_p} e^{-\frac{j2\pi}{M}mk_p} \delta[k-nM] \delta[l-mN].
    \end{equation*}

    Therefore,
    \begin{align*}
        \widehat{\mathbf{h}}_{\mathrm{eff}}[k', l'] &=\!\sum_{k, l \in \mathbb{Z}}\!\mathbf{h}_{\mathrm{eff}}[k, l] \mathbf{A}_{\mathbf{x}_{(k_p, l_p)}^{(\mathrm{p})}}[k'-k, l'-l] e^{\frac{j2\pi}{MN}l(k'-k)} \nonumber \\
        &= \mathbf{h}_{\mathrm{eff}}[k', l'] + \sum_{(n,m)\neq(0,0)} e^{\frac{j2\pi}{N}n(l_p+l')} e^{-\frac{j2\pi}{M}mk_p} \nonumber \\ &~~~~~~~~~~~~~~~~~~~~~~~~~~~\times \mathbf{h}_{\mathrm{eff}}[k'-nM, l'-mN].
    \end{align*}

    Therefore, $\widehat{\mathbf{h}}_{\mathrm{eff}}[k', l'] = \mathbf{h}_{\mathrm{eff}}[k', l']$ if the channel support satisfies the condition in the Theorem statement.

    For spread carrier-based Zak-OTFS, from Theorem~\ref{thm:frft_amb}:
    \begin{equation*}
        \mathbf{A}_{\mathbf{x}_{(k_p, l_p)}^{(\mathrm{c})}}[k, l]\!=\!e^{\frac{j2\pi}{MN} (-Ak^2+lk+CB^{-2}_{{}_{MN}} (2Ak-l)^2)} \mathbf{A}_{\mathbf{x}_{(k_p, l_p)}^{(\mathrm{p})}}[\bar{k}, \bar{l}],
    \end{equation*}
    where $\bar{k} = -B^{-1}_{{}_{MN}} (2Ak-l)$, $\bar{l} = 2CB^{-1}_{{}_{MN}} (2Ak-l) - Bk$.

    On substituting the expression for $\mathbf{A}_{\mathbf{x}_{(k_p, l_p)}^{(\mathrm{p})}}[k, l]$,
    \begin{align*}
        \mathbf{A}_{\mathbf{x}_{(k_p, l_p)}^{(\mathrm{c})}}[k, l]\!&=\!e^{\frac{j2\pi}{MN} (-Ak^2+lk+CB^{-2}_{{}_{MN}} (2Ak-l)^2)}\!\sum_{n,m\in\mathbb{Z}} e^{\frac{j2\pi}{N}nl_p} \nonumber \\
        &~~~~~~~~~~~~~~~\times e^{-\frac{j2\pi}{M}mk_p} \delta[\bar{k}-nM] \delta[\bar{l}-mN].
    \end{align*}

    The delta functions in terms of $(\bar{k},\bar{l})$ may be equivalently expressed in terms of $(k,l)$ by solving the $2 \times 2$ linear equations $\bar{k} = -B^{-1}_{{}_{MN}} (2Ak-l) = nM$, $\bar{l} = 2CB^{-1}_{{}_{MN}} (2Ak-l) - Bk = mN$ for $(k,l)$. The solution gives:
    \begin{align*}
        \mathbf{A}_{\mathbf{x}_{(k_p, l_p)}^{(\mathrm{c})}}[k, l]\!&=\!e^{\frac{j2\pi}{MN} (-Ak^2+lk+CB^{-2}_{{}_{MN}} (2Ak-l)^2)}\!\sum_{n,m\in\mathbb{Z}} e^{\frac{j2\pi}{N}nl_p} \nonumber \\
        &~~~~~~~~~~~~~~\times e^{-\frac{j2\pi}{M}mk_p} \delta[k-k'_{n,m}] \delta[l-l'_{n,m}],
    \end{align*}
    where $(k'_{n,m},l'_{n,m})$ are as defined in the Theorem statement. Substituting into the expression for $\widehat{\mathbf{h}}_{\mathrm{eff}}[k', l']$ gives the desired condition for $\widehat{\mathbf{h}}_{\mathrm{eff}}[k', l'] = \mathbf{h}_{\mathrm{eff}}[k', l']$ to hold.
\end{IEEEproof}

\textcolor{black}{In practice, the GDAFT parameters $A,B,C$ must be chosen such that the crystallization condition in Lemma~\ref{lmm:heff_cryst} holds. We illustrate the choice of parameters $A,B,C$ via an example.}

\begin{figure}
\centering
\begin{subfigure}{0.49\columnwidth}
    \includegraphics[width=\textwidth]{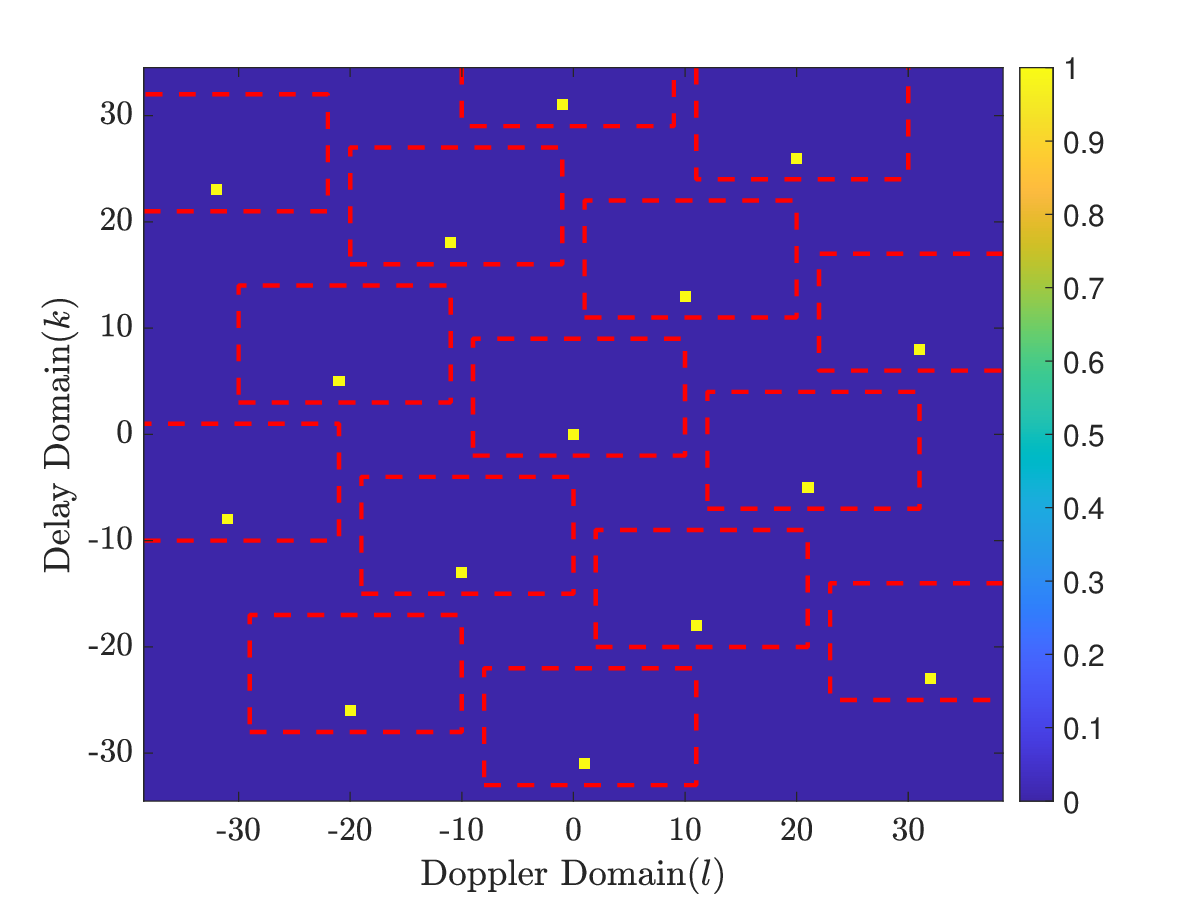}
    %\vspace*{-0.1in}
\caption{$A = 3, B = 5, C = 7$.}
    \label{fig:GDAFT_ex1}
\end{subfigure}
\begin{subfigure}{0.49\columnwidth}
    \includegraphics[width=\textwidth]{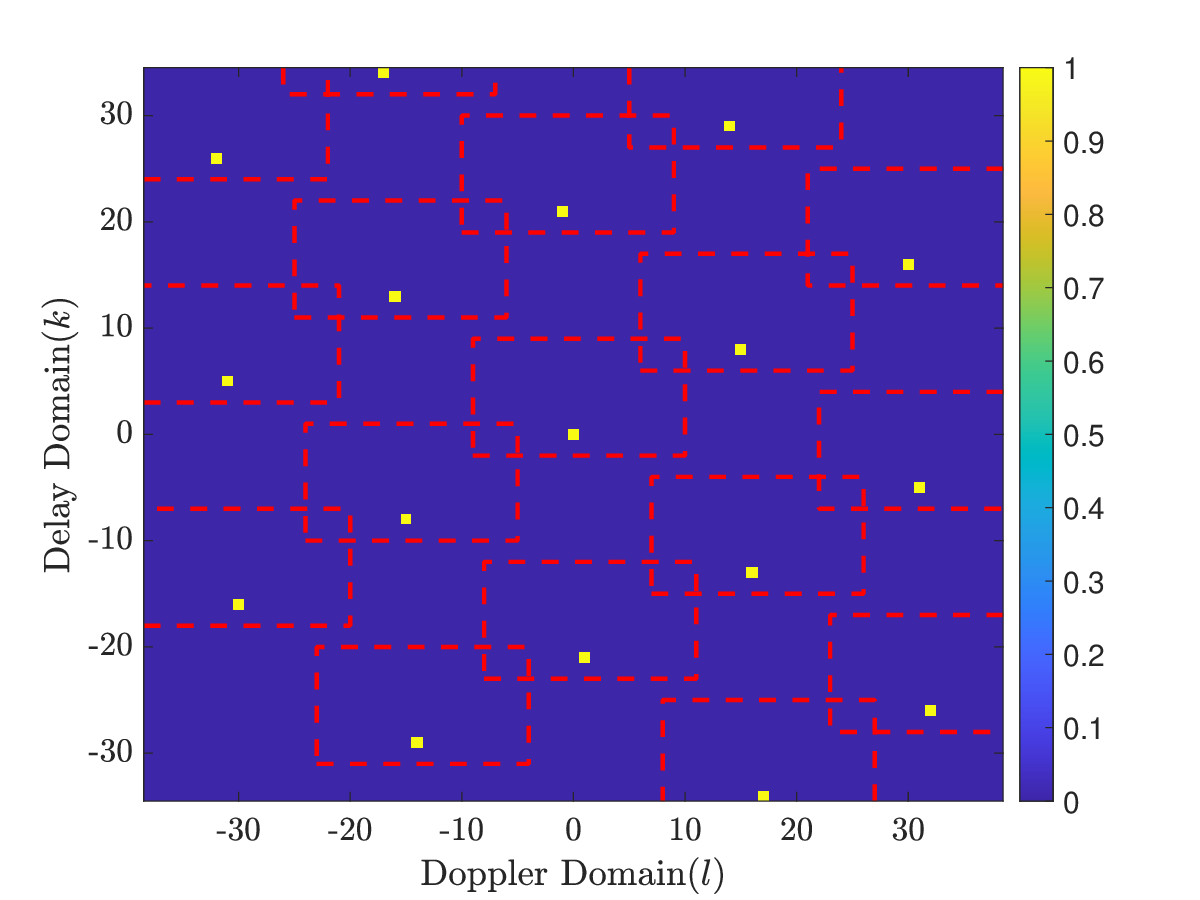}
    %\vspace*{-0.1in}
\caption{$A = 2, B = 5, C = 7$.}
    \label{fig:GDAFT_ex2}
\end{subfigure}
%\vspace*{-0.1in}
\caption{\textcolor{black}{Example illustrating choice of GDAFT parameters.}}
\vspace{-5mm}
    \label{fig:GDAFT_ex}
\end{figure}

\textcolor{black}{\textit{Example:} Consider $M = 17, N = 19$ and channel support $\mathcal{S} = [k_{\min},k_{\max}] \times [l_{\min},l_{\max}]$ with\footnote{\textcolor{black}{$k_{\min} < 0$ to accommodate delay spread due to the pulse shaping filter.}} $k_{\min} = -2$, $k_{\max} = 8$, $l_{\min} = -9$, $l_{\max} = 9$. Fig.~\ref{fig:GDAFT_ex} illustrates the support of the estimated effective channel $\widehat{\mathbf{h}}_{\mathrm{eff}}[k, l]$ in~\eqref{eq:chest2} for different choices of parameters $A,B,C$, given by translates of the channel support $\mathcal{S}$ on locations $(k'_{n,m},l'_{n,m})$ per Lemma~\ref{lmm:heff_cryst} and shown as dashed red rectangles. For $A = 3, B = 5, C = 7$, the rectangles don't overlap, and hence the effective channel can be perfectly estimated for all $(k,l) \in \mathcal{S}$. However, the rectangles overlap when $A = 2, B = 5, C = 7$, and accurate channel estimation is not possible in this case.} 

\textcolor{black}{In practice, grid search is performed to determine $A,B,C$. Note that prior knowledge of channel support $\mathcal{S}$ \emph{does not} need to be exact, and estimates of $k_{\min},k_{\max},l_{\min},l_{\max}$, e.g., from the maximum expected delay to the cell edge and maximum speed of objects, can be used to approximate $\mathcal{S}$.}

\subsection{Data Detection (Data Stage)}
\label{subsec:cazac_proc_datadetecn}

In the data stage, the receiver performs data detection via the minimum mean squared error (MMSE) estimator~\cite{Tse2005}:
\begin{equation}
    \label{eq:detecn1}
    \widehat{\mathbf{x}} = (\widehat{\mathbf{H}}^{\mathsf{H}}\widehat{\mathbf{H}}+\sigma^2\mathbf{I})^{-1} \widehat{\mathbf{H}}^{\mathsf{H}} \mathbf{y}.
\end{equation}

\section{Numerical Results}
\label{sec:results}

We simulate the system model in~\eqref{eq:sys_model3} for both pulsone-based and spread carrier-based Zak-OTFS. The effective channel $\mathbf{h}_{\mathrm{eff}}[k,l]$ in~\eqref{eq:sys_model1} is generated by sampling the continuous effective channel as~\cite{bitspaper1,bitspaper2,Aug2024paper}:
\begin{align}
    \label{eq:heff1}
    \mathbf{h}_{\mathrm{eff}}[k,l] &= \mathbf{h}_{\mathrm{eff}}\bigg(\tau = \frac{k\tau_p}{M},\nu = \frac{l\nu_p}{N}\bigg), \nonumber \\
    \mathbf{h}_{\mathrm{eff}}(\tau,\nu) &= \mathbf{w}_{{}_\mathrm{RX}}(\tau,\nu) *_\sigma \mathbf{h}_{\mathrm{phy}}(\tau,\nu) *_\sigma \mathbf{w}_{{}_\mathrm{TX}}(\tau,\nu),
\end{align}
where $*_\sigma$ denotes twisted convolution\footnote{$a(\tau,\nu)*_\sigma b(\tau,\nu) = \iint a(\tau',\nu') b(\tau-\tau',\nu-\nu') e^{j2\pi\nu'(\tau-\tau')} d\tau' d\nu'$.}. We assume sinc transmit pulse shaping and matched filtering at the receiver~\cite{Aug2024paper}:
\begin{align}
    \label{eq:pulseshaping1}
    \mathbf{w}_{{}_\mathrm{TX}}(\tau,\nu) &= \sqrt{BWT}~\mathrm{sinc}(BW\tau)~\mathrm{sinc}(T\nu), \nonumber \\
    \mathbf{w}_{{}_\mathrm{RX}}(\tau,\nu) &= e^{j2\pi\nu\tau} \mathbf{w}_{{}_\mathrm{TX}}^*(-\tau,-\nu),
\end{align}
where $BW = M\nu_p,T=N\tau_p$. The physical channel in~\eqref{eq:heff1}, $\mathbf{h}_{\mathrm{phy}}(\tau,\nu) = \sum_{i=1}^{P} h_i \delta(\tau-\tau_i) \delta(\nu-\nu_i)$, corresponds to a 3GPP-compliant $P=6$ path Vehicular-A (Veh-A) channel model~\cite{veh_a} with significant mobility and delay spread, whose power-delay profile is shown in Table~\ref{tab:veh_a}. The Doppler of each path is simulated as $\nu_i = \nu_{\max}\cos(\theta_i)$, with $\theta_i$ uniformly distributed in $[-\pi, \pi)$ and $\nu_{\max} = 815$ Hz denoting the maximum channel Doppler spread\footnote{\textcolor{black}{Our channel model considers \textit{fractional} delay and Doppler shifts, which is representative of real propagation environments. The path delays in Table~\ref{tab:veh_a} are non-integer multiples of the delay resolution $\nicefrac{1}{BW}$. The Doppler shifts $\nu_i = \nu_{\max}\cos(\theta_i)$ are also non-integer multiples of the Doppler resolution $\nicefrac{1}{T}$ since $\cos(\theta_i)$ is continuous valued.}}. \textcolor{black}{We simulate both narrowband and wideband system regimes with parameters: $M=17, N=19$, $BW = 0.51$ MHz (narrowband) and $M=83, N=13$, $BW = 19.9$ MHz (wideband).} For the spread carrier basis, we choose the GDAFT parameters $A = 3, B = 5, C = 7$ such that the crystallization condition in Lemma~\ref{lmm:heff_cryst} holds. The transmitted data symbols $\mathbf{x}_{d}$ in~\eqref{eq:sys_model3} are chosen from an uncoded $4$-QAM constellation.

\begin{table}[!t]
    \centering
    \caption{Power-delay profile of Veh-A channel model}
    \begin{tabular}{|c|c|c|c|c|c|c|}
         \hline
         Path index $i$ & 1 & 2 & 3 & 4 & 5 & 6 \\
         \hline
         Delay $\tau_i (\mu s)$ & 0 & 0.31 & 0.71 & 1.09 & 1.73 & 2.51 \\
         \hline
         Relative power (dB) & 0 & -1 & -9 & -10 & -15 & -20 \\
         \hline
    \end{tabular}
    \label{tab:veh_a}
\end{table}

% \begin{figure}
% \centering
% \begin{subfigure}{0.49\columnwidth}
%     \includegraphics[width=\textwidth]{tdwaveform_pulsone.eps}
%     %\vspace*{-0.1in}
% \caption{Pulsone basis element.}
%     \label{fig:doublebpaths_1}
% \end{subfigure}
% \begin{subfigure}{0.49\columnwidth}
%     \includegraphics[width=\textwidth]{tdwaveform_cazac.eps}
%     %\vspace*{-0.1in}
% \caption{Spread carrier basis element.}
%     \label{fig:doublebpaths_2}
% \end{subfigure}
% %\vspace*{-0.1in}
% \caption{Time-domain waveforms of pulsone and spread carrier basis elements corresponding to $(k_0,l_0) = ((M+1)/2,(N+1)/2)$. The former has a PAPR of $15.1$ dB, whereas the latter has a PAPR of $3.58$ dB. Note that only the real part of a portion of the entire waveform is shown.}% (per~\eqref{eq:zakotfs1} and Theorem~\ref{thm:frft_pp_tdcazac} respectively) with a sinc transmit pulse shaping filter. Waveform parameters are $(k_0,l_0) = ((M+1)/2,(N+1)/2)$, $M = 17, N = 19$, $\nu_{p} = 30$ kHz, sampling frequency $f_s = M\nu_p$.}
% %\vspace*{-0.1in}
%     \label{fig:tdwvf}
% \end{figure}

\begin{figure}[!t]
    \centering
    \includegraphics[width=0.8\linewidth]{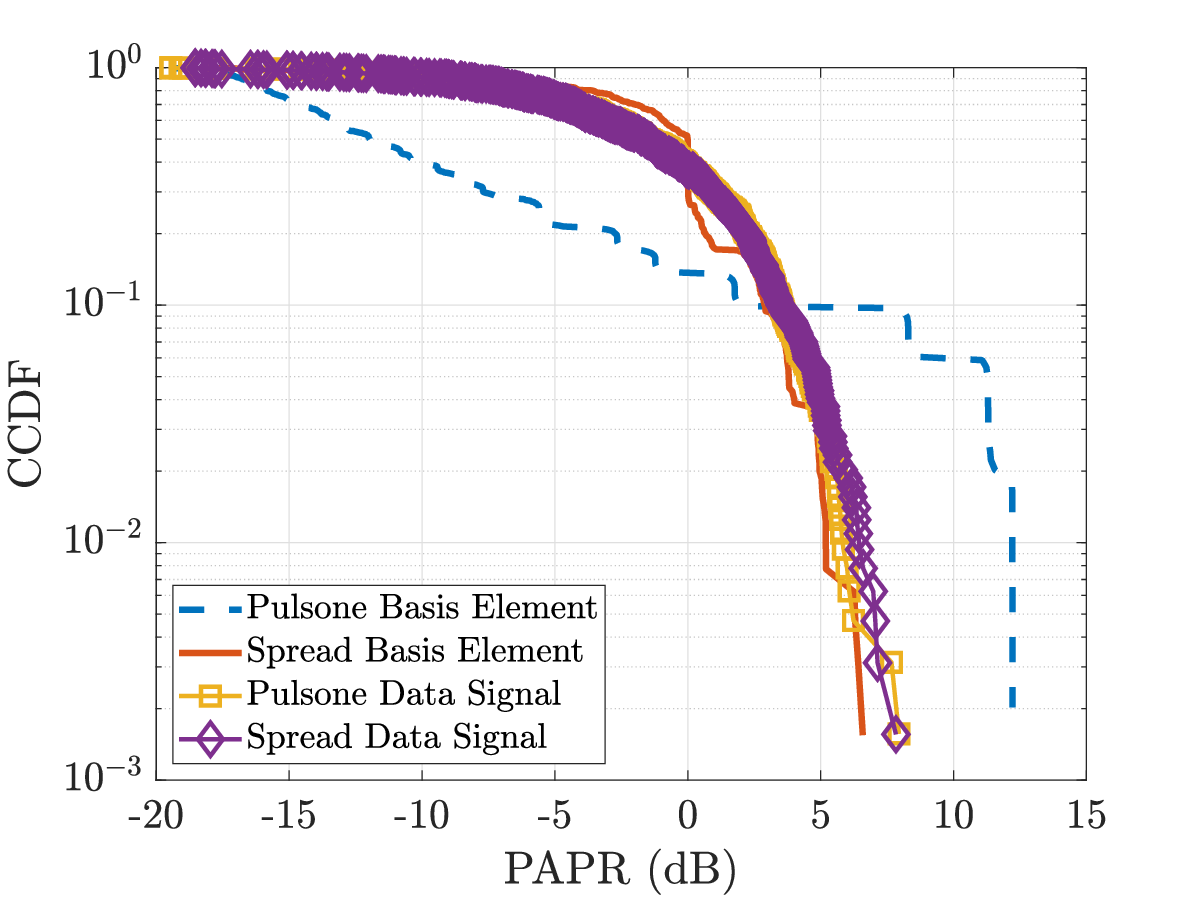}
    \caption{\textcolor{black}{PAPR of pulsone- and spread carrier-based Zak-OTFS.}}
    \label{fig:papr_ccdf}
    \vspace{-2mm}
\end{figure}

\subsection{Peak-to-Average Power Ratio (PAPR) Comparison}
\label{subsec:results_papr}

\textcolor{black}{Fig.~\ref{fig:papr_ccdf} plots the complementary cumulative distribution function (CCDF) of the PAPR for pulsone- and spread carrier-based Zak-OTFS for system parameters: $M=17, N=19$, $BW = 0.51$ MHz and an oversampling factor of $4$. Due to the CAZAC property of the spread carrier basis (see Theorem~\ref{thm:frft_pp_tdcazac}), each spread carrier basis element has $5.6$ dB lower PAPR than each pulsone basis element. When modulated by information symbols, the spread carrier data signal has $7.83$ dB PAPR compared to $7.95$ dB PAPR for the pulsone data signal.}

% time-domain waveforms of the pulsone and spread carrier basis elements with a sinc transmit pulse shaping filter. Due to the constant amplitude property of CAZACs, the latter has $11.5$ dB lower PAPR than the former. %In comparison, the PAPR of a single OFDM subcarrier is $22.2$ dB.

\subsection{Channel Estimation Performance}
\label{subsec:results_channelest}

\begin{figure}[!t]
    \centering
    \includegraphics[width=0.8\linewidth]{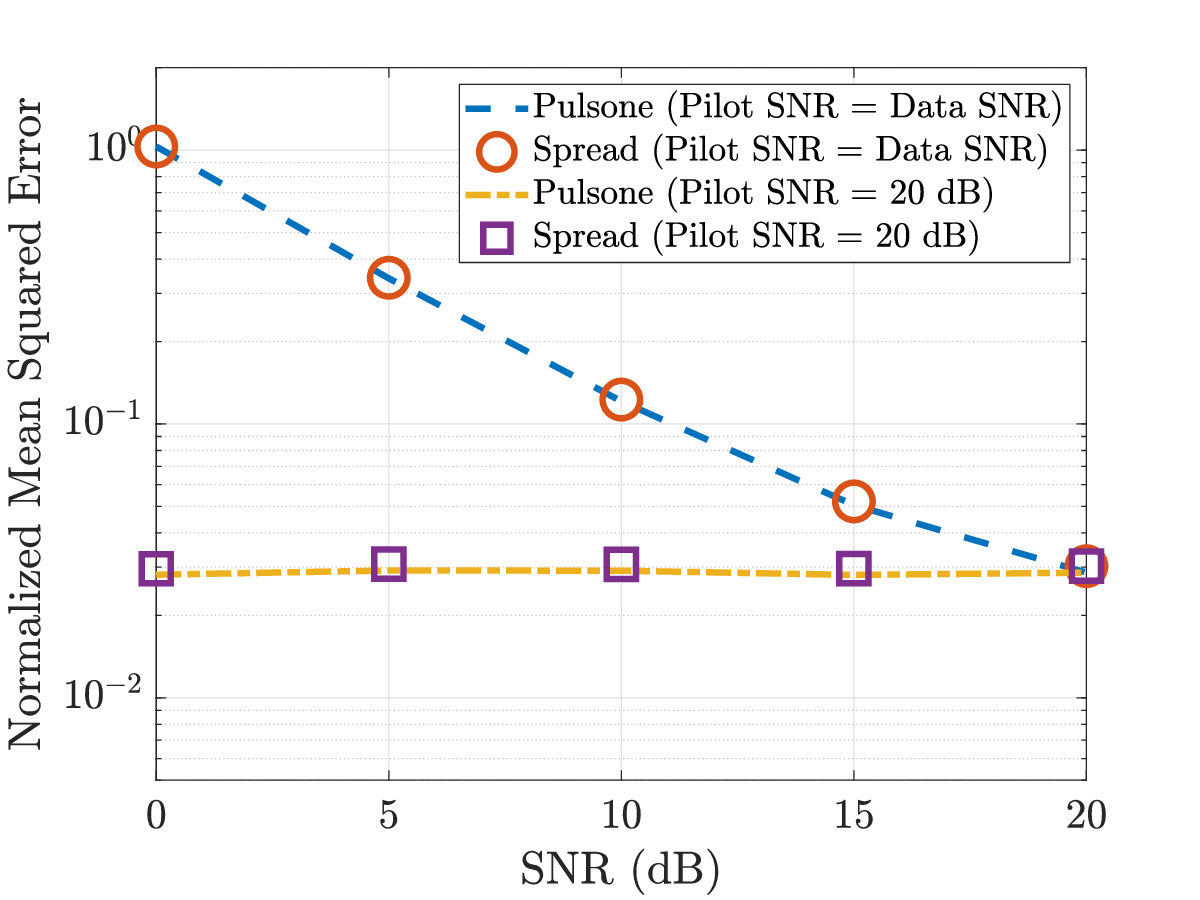}
    \caption{\textcolor{black}{Normalized mean squared error in channel estimation is the same for pulsone- and spread carrier-based Zak-OTFS.}}%, assuming sinc pulse shaping, a $6$-path Vehicular-A channel as per Table~\ref{tab:veh_a}, and parameters $M = 17$, $N = 19$, $\nu_p = 30$ kHz.}
    \label{fig:nmse}
    \vspace{-2mm}
\end{figure}

Fig.~\ref{fig:nmse} plots the normalized mean squared error (NMSE) in channel estimation via~\eqref{eq:chest2} for pulsone- and spread carrier-based Zak-OTFS \textcolor{black}{for system parameters: $M=17, N=19$, $BW = 0.51$ MHz.} We consider two scenarios: (i) equal signal-to-noise ratio (SNR) in the pilot and data stages\footnote{\textcolor{black}{The transmit power remains same for pilot and data frames in this case.}} in~\eqref{eq:sys_model3}, and (ii) \textcolor{black}{with $20$ dB pilot SNR. The NMSE of the former reduces with increasing pilot SNR and approaches the NMSE of the latter at $20$ dB SNR.} In both scenarios, the NMSEs match for pulsones and spread carriers, i.e., the systems are equivalent. We note that for data detection, we only require channel estimates to be accurate enough (but not perfect).

\vspace{-2mm}

\subsection{Data Detection Performance}
\label{subsec:results_ber}

\begin{figure}[!t]
    \centering
    \includegraphics[width=0.8\linewidth]{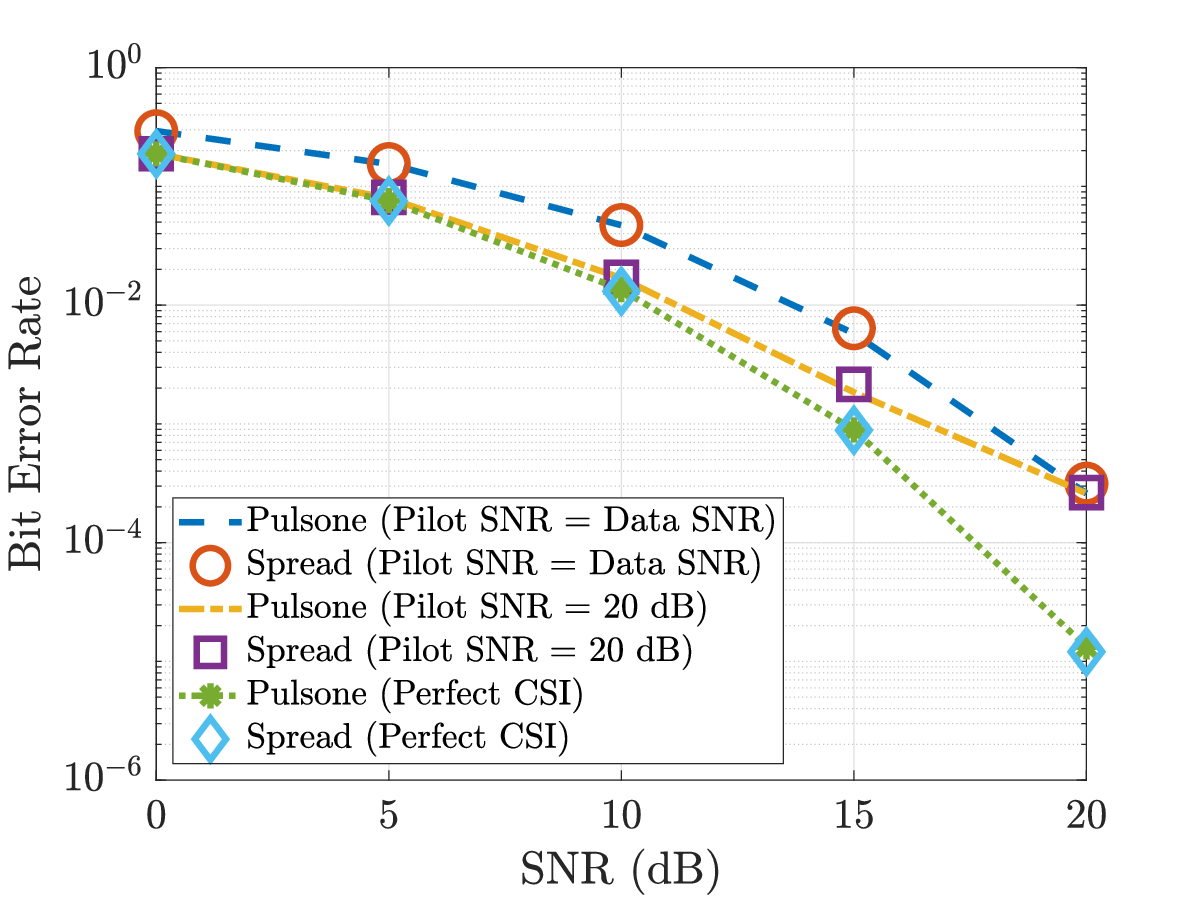}
    \caption{\textcolor{black}{Bit error rate curves showing the equivalence of data detection with pulsone- and spread carrier-based Zak-OTFS.}}%, assuming sinc pulse shaping, a $6$-path Vehicular-A channel as per Table~\ref{tab:veh_a}, and parameters $M = 17$, $N = 19$, $\nu_p = 30$ kHz.}
    \label{fig:ber}
    \vspace{-2mm}
\end{figure}

Fig.~\ref{fig:ber} plots the bit error rate (BER) curves for pulsone- and spread carrier-based Zak-OTFS \textcolor{black}{for system parameters: $M=17, N=19$, $BW = 0.51$ MHz} and three scenarios: (i) perfect channel knowledge, (ii) equal pilot and data SNR, and (iii) $20$ dB pilot SNR. \textcolor{black}{The BER with $20$ dB pilot SNR diverges from the perfect channel knowledge curve at higher SNRs, matching the equal pilot and data SNR curve at $20$ dB SNR and saturating beyond $20$ dB SNR.} In all three scenarios, the BER curves match for pulsones and spread carriers, thus demonstrating the equivalence of pulsone-based and the proposed spread carrier-based Zak-OTFS systems.

% The BER with $40$ dB pilot SNR is equivalent to that with perfect channel knowledge. The BER curve for equal pilot and data SNR approaches the other curves with increasing SNR. In all cases, the BERs match for pulsones and spread carriers.%is the same for both bases, thus demonstrating their equivalence.

\subsection{\textcolor{black}{Comparison with Existing Approaches}}
\label{subsec:results_comparison}

\begin{figure}[!t]
    \centering
    \includegraphics[width=0.8\linewidth]{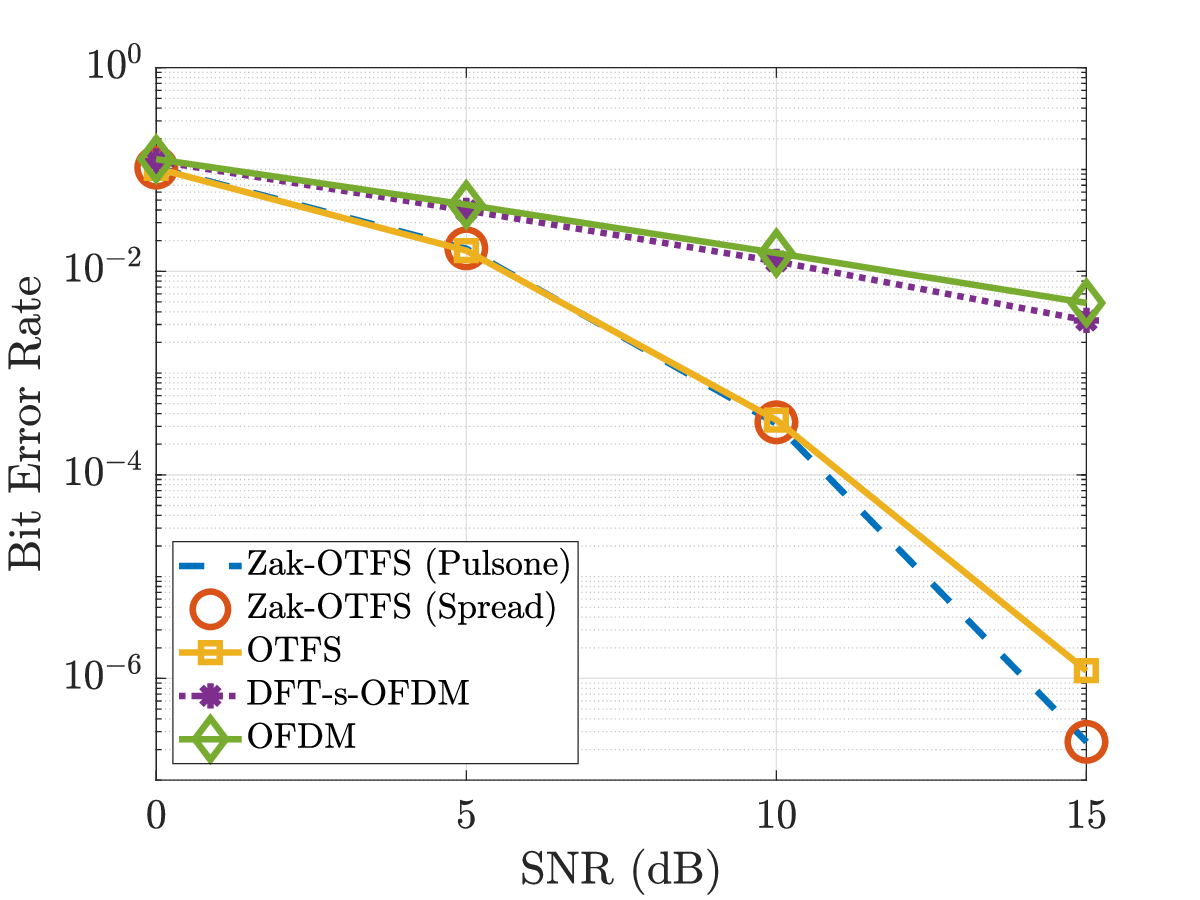}
    \caption{\textcolor{black}{Bit error rate curves comparing the performance of the proposed spread Zak-OTFS scheme with existing approaches.}}
    \label{fig:comparison}
    \vspace{-2mm}
\end{figure}

\textcolor{black}{Fig.~\ref{fig:comparison} plots the BER curves for system parameters: $M=17, N=19$, $BW = 0.51$ MHz assuming perfect channel knowledge across five systems: pulsone-based Zak-OTFS, spread carrier-based Zak-OTFS, OTFS, OFDM and DFT-spread-OFDM. To ensure frame consistency between the latter three and Zak-OTFS, we choose the subcarrier spacing as $\nu_p$, number of subcarriers as $M$, number of symbols per frame as $N$. The cyclic prefix length is $\lceil BW\tau_{\max} \rceil \nicefrac{1}{BW}$ for maximum delay spread $\tau_{\max}$ of the channel in Table~\ref{tab:veh_a}. Since DFT-spread-OFDM only encodes information on a fraction of the total number of available subcarriers, say $L$ out of the total $M$, for fair comparison we ensure that information is modulated on only $LN$ (randomly chosen) symbols out of the total $MN$ symbols per frame in Zak-OTFS, OTFS and OFDM. Fig.~\ref{fig:comparison} corresponds to the choice $L = 9$, for a spectral efficiency of $\nicefrac{L}{M} = \nicefrac{9}{17}$. The BER curves demonstrate the value of delay-Doppler signaling with non-fading modulations like Zak-OTFS in doubly-selective channels~\cite{bitspaper1,bitspaper2}. The OFDM-based time-frequency modulations have poor BER due to channel fading. OTFS yields significantly better performance than both OFDM systems, and the performance improves further at $15$ dB SNR with Zak-OTFS due to the greater predictability of the Zak-OTFS I/O relation. Moreover, the BER performance of pulsone- and spread carrier-based Zak-OTFS is the same.} %and that the Zak-OTFS BER curves are the lowest among all considered systems. Specifically, Zak-OTFS outperforms OTFS at $15$ dB SNR and outperforms OFDM \& DFT-spread-OFDM across all SNR values.

\subsection{\textcolor{black}{Narrowband vs Wideband Performance}}
\label{subsec:results_bw}

\begin{figure}[!t]
    \centering
    \includegraphics[width=0.8\linewidth]{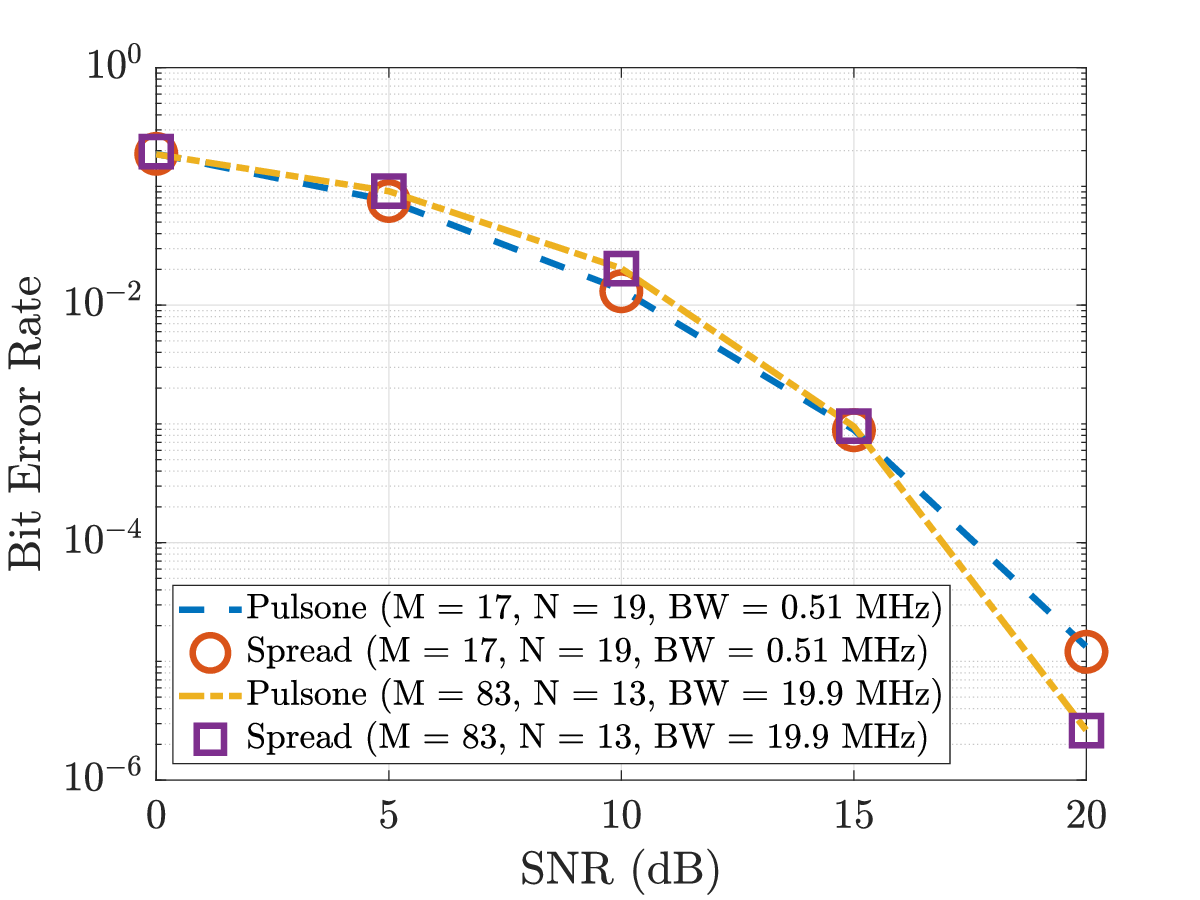}
    \caption{\textcolor{black}{Bit error rate curves showing similar data detection performance for narrowband and wideband systems.}}%, assuming sinc pulse shaping, a $6$-path Vehicular-A channel as per Table~\ref{tab:veh_a}, and parameters $M = 17$, $N = 19$, $\nu_p = 30$ kHz.}
    \label{fig:ber_20mhz}
    \vspace{-2mm}
\end{figure}

\textcolor{black}{Fig.~\ref{fig:ber_20mhz} plots the BER curves for pulsone- and spread carrier-based Zak-OTFS for both narrowband and wideband system parameters assuming perfect channel knowledge. The wideband system parameters are chosen to approximate 5G-NR numerology: $N = 13$ (5G-NR has $12-14$ symbols per slot), $\nu_p = 240$ kHz (5G-NR subcarrier carrier spacing) and $M = 83$ for $BW = 19.9$ MHz. The BER performance remains similar for narrowband and wideband systems; hence, our proposed method can be utilized in both regimes.}

% \begin{figure}[!t]
%     \centering
%     \includegraphics[width=0.8\linewidth]{pulsone_vs_cazac_ber_noheff_perfectcsi_A3_B5_C7_M17_N19.eps}
%     \caption{Bit error rate curves showing equivalence of realizing Zak-OTFS with pulsones and CAZACs, assuming perfect channel knowledge at the receiver, no pulse shaping and a $6$-path Vehicular-A channel with integer delay and Doppler when sampled on a DD grid with $M = 17$, $N = 19$, $\nu_p = 30$ kHz.}
%     \label{fig:ber_noheff_perfectcsi}
% \end{figure}

\section{Conclusion}
\label{sec:conclusion}

\textcolor{black}{In this paper, we proposed a spread carrier implementation of Zak-OTFS by transforming high-PAPR pulsones to low-PAPR spread carrier waveforms via a generalization of the discrete affine Fourier transform. We showed that the proposed system achieves full spectral efficiency like Zak-OTFS with $5.6$ dB lower PAPR per basis element, and outperforms competing methods based on OFDM and OTFS. Future work will explore applications of the proposed method to security, radar sensing, and integrated sensing \& communication.}

% In this paper, we proposed a spread spectrum communication system using Zak-OTFS by transforming high-PAPR pulsones to low-PAPR CAZAC sequences via a unitary transformation. We demonstrated the equivalence of both systems via bit error rate simulations with a six-path Vehicular-A channel model. Future work will further explore the application of the proposed spread spectrum Zak-OTFS in secure communication, radar sensing, and integrated sensing \& communication.

%In this paper, we proposed an equivalent implementation of Zak-OTFS using CAZAC sequences instead of pulsones. The benefit of the former over the latter is easier standard adoption and lower peak-to-average power ratio. We demonstrated the equivalence of both implementations via bit error rate simulations with a six-path Vehicular-A channel model. Future work will investigate the utility of the proposed CAZAC-based Zak-OTFS implementation in radar sensing, integrated sensing and communication, and grant-free multiple access.

\bibliographystyle{IEEEtran}
\bibliography{references}

\end{document}